\documentclass[%
 reprint,
superscriptaddress,
 amsmath,amssymb,
 aps,
]{revtex4-2}

\usepackage[final]{changes}
\usepackage{graphicx}
\usepackage{dcolumn}
\usepackage{bm}
\usepackage[
   colorlinks=true,
   citecolor=blue,
   linkcolor=blue,
   urlcolor=blue,
]{hyperref}

\usepackage{xcolor}
\usepackage{soul}

\usepackage{amsmath}
\usepackage{upgreek}
\usepackage{tabularx}
\usepackage{braket}
\usepackage{siunitx}   
\usepackage{braket}

\usepackage{placeins}

\def\orcid#1{\kern .08em\href{https://orcid.org/#1}{\includegraphics[keepaspectratio,width=0.7em]{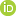}}}

\begin{document}

\preprint{APS/123-QED}

\title{Raman Transitions in Collinear Laser Spectroscopy for High-Precision Frequency Measurements} 
\author{J. Spahn\orcid{0009-0007-8354-4896}}
    \affiliation{Institut für Kernphysik, Technische Universität Darmstadt, 64289 Darmstadt, Germany}
    \email{jspahn@ikp.tu-darmstadt.de}
\author{H. Bodnar\orcid{0009-0005-3056-3124}}
    \affiliation{Institut für Kernphysik, Technische Universität Darmstadt, 64289 Darmstadt, Germany}
\author{Kristian König\orcid{0000-0001-9415-3208}}
    \affiliation{Institut für Kernphysik, Technische Universität Darmstadt, 64289 Darmstadt, Germany}%
    \affiliation{Helmholtz Research Academy Hesse for FAIR, GSI Helmholtzzentrum für Schwerionenforschung, 64291 Darmstadt, Germany}
\author{W. Nörtershäuser\orcid{0000-0001-7432-3687}}
    \affiliation{Institut für Kernphysik, Technische Universität Darmstadt, 64289 Darmstadt, Germany}%
    \affiliation{Helmholtz Research Academy Hesse for FAIR, GSI Helmholtzzentrum für Schwerionenforschung, 64291 Darmstadt, Germany}
\author{R. Van Duyse\orcid{0009-0003-6034-2184}}
    \affiliation{Institute for Nuclear and Radiation Physics, KU Leuven, 3000 Leuven, Belgium}


\date{\today}

\newcommand{\Sone}{$\mathrm{S}_{1/2}$}
\newcommand{\Pthree}{$\mathrm{P}_{3/2}$}
\newcommand{\Dthree}{$\mathrm{D}_{3/2}$}
\newcommand{\Dfive}{$\mathrm{D}_{5/2}$}

\newcommand{\SonePthree}{$\mathrm{S}_{1/2}\rightarrow \mathrm{P}_{3/2}$}
\newcommand{\DthreePthree}{$\mathrm{D}_{3/2}\rightarrow \mathrm{P}_{3/2}$}
\newcommand{\DfivePthree}{$\mathrm{D}_{5/2}\rightarrow \mathrm{P}_{3/2}$}

\newcommand{\SoneDthree}{$\mathrm{S}_{1/2}\rightarrow \mathrm{D}_{3/2}$}
\newcommand{\SoneDfive}{$\mathrm{S}_{1/2}\rightarrow \mathrm{D}_{5/2}$}
\newcommand{\DthreeDfive}{$\mathrm{D}_{3/2}\rightarrow \mathrm{D}_{5/2}$}

\begin{abstract}

We demonstrate stimulated Raman spectroscopy on dipole-forbidden fine-structure transitions in a fast ion beam.
It is applied to the $\mathrm{S}_{1/2}\rightarrow\mathrm{D}_{5/2}$ clock transition in $^{88}\mathrm{Sr}^{+}$ and combined with a subsequent E1 transition to allow for its detection by standard fluorescence detection. Building on this, the $\mathrm{D}_{3/2}\rightarrow\mathrm{D}_{5/2}$ fine-structure splitting is measured directly using a Doppler-free scheme employing two subsequent Raman transitions. These measurements demonstrate a path toward precision collinear laser spectroscopy of narrow transitions on short-lived isotopes beyond the linewidth limit of conventional electric-dipole spectroscopy, which can be used to probe fundamental aspects of atomic and nuclear physics.

\end{abstract}

\maketitle


\section{\label{sec:Introduction}Introduction}

\subsection{\label{ssec:Motivation}Motivation}
Collinear laser spectroscopy (CLS) is a technique in use for about 50 years to study ionic and atomic transitions \cite{Wing.1976,Meier.1977,anton1978collinear}. Doppler compression of the ion-beam velocity distribution \cite{Kaufman.1976} via electrostatic acceleration with several $10\,$kV provides narrow linewidth, typically of the order of a few 10\,MHz, which is finally limited by the natural linewidth of the usually studied electric-dipole (E1) transitions. Combined with the quasi-simultaneous collinear-anticollinear approach \cite{Borghs.1981} and frequency-comb metrology \cite{Nortershauser.2009}, it has reached a level of precision \cite{ImgrBa} comparable with the most accurate frequency measurements of E1 transitions in ion traps \cite{Gebert.2015}. In Ca$^+$ ions, this has, for example, been used to test state-of-the-art atomic-structure calculations, with two approaches being prone to different systematics \cite{Shi.2016,PaMuCaPhysRevResearch}. 
Additionally, being a fast method with transport and interrogation times in the order of $\upmu$s, it is especially well suited to investigate short-lived isotopes \cite{Schinzler.1978,Blaum.2013,GarciaRuiz.2020,yang2023laser,Koszorus2024CLSReview} but also short-lived metastable electronic states in atoms and ions \cite{Thompson.1998,Myers.1999,Imgr12CPhysRevLett,Muller.2025,Koenig2026C14}. The comparably high resolution has also been exploited to study fundamental aspects of atomic, nuclear and particle physics, like tests of relativistic atomic structure and QED calculations on stable and radioactive isotopes \cite{Riis.1994,Wendt.1984,Krieger2017Be,ImgrBa,Mueller2026Splitting}, tests of special relativity \cite{Poulsen.1983}, or photo-detachment and spectroscopy of negative ions \cite{Berzinsh.1995,Warring.2009,Leimbach.2020}. A final example is the application of CLS for high-voltage metrology \cite{Poulsen.1988,kramer2018high}. 

Already in its early years, CLS was combined with several other techniques used in atomic and molecular spectroscopy to reduce the linewidth and, thus, enhance accuracy. For instance, three-level saturation spectroscopy in V- and $\Lambda$-spectroscopy was demonstrated \cite{Poulsen.1983} and applied for tests of time dilation in special relativity (Ives-Stilwell experiment) in fast beams \cite{Riis.1988}. Later, it allowed the application to relativistic ion beams in storage rings for which the Doppler compression does not hold \cite{Saathoff.2003b,Reinhardt.2007b,Botermann.2014}. CLS was also used to study transient effects in photon-ion interactions \cite{borghs_transient_1981,Wannstrom.1988,jovanovic_modeling_2023}, and Ramsey interference fringes were observed using fast beam Doppler switching by high voltages \cite{silverans_observation_1981}.  

All these efforts have in common that they still apply E1 transitions and will finally run into the natural-linewidth limit of the order of a few MHz in the best cases. Several applications that are currently discussed would profit from studying dipole-forbidden transitions offering narrower lines. The intercombination transition in In$^+$ ions has for example been studied as a possibility to improve high-voltage measurements \cite{Koenig2020In}. King plots are used to search for new bosons \cite{Frugiuele.2017,Berengut.2018,Door.2025,Fuchs.2025,Wilzewski.2025} and could be expanded to long chains of short-lived isotopes if CLS could be performed on forbidden transitions. It might also support the extraction of higher-order nuclear moments from isotope shift and hyperfine structure, like magnetic octupole moments \cite{deGroote2022Sc45}, the fourth-order radial moment of the charge density \cite{Papoulia.2016,PGR4thRadialMoment}, or the Zemach radius \cite{Sun.2023}.
The use of CLS for studying E1-forbidden transitions is sparse. M1 transitions have been studied for QED tests in the strong magnetic fields in highly charged heavy ions \cite{Klaft.1994,Ullmann2017,Horst.2025}, where the long lifetime of the excited state is counteracted by long observation times in storage rings. 
The RF-optical double resonance technique has been adopted to CLS for high-precision octupole measurements, where optical E1 transitions and collinear resonance ionization spectroscopy (CRIS) are used to detect M1 transitions driven by RF fields in atoms \cite{deGroote2022Sc45}. Similarly, CRIS was combined with first-order Doppler-free two-photon spectroscopy on Rb atoms \cite{Yang.2026}.

Another approach to overcoming E1-linewidth limitations is Raman transitions, as their intrinsic linewidth is about two orders of magnitude smaller \cite{dunning2015coherent, Solaro2018}. In the past, to our knowledge, Raman transitions have only once been used in CLS for hyperfine structure studies by driving the Raman transition using a single laser and its radio-frequency-shifted sideband \cite{dinneen1991stimulated}, providing a very high degree of coherence between the two laser fields but limiting the method to transitions between hyperfine states of the same fine structure level.
Using two separate lasers allows this method to be expanded to transitions between different states of equal parity that can differ in the main quantum number, which are of interest in state-of-the-art atomic and molecular physics, and are being studied, e.g., in ion traps \cite{Kurth.1995,Leibfried.2001,Yamazaki.2008}.

For CLS, this class of Raman transitions has so far only been proposed \cite{neumann2020raman} and experimentally demonstrated as a velocity filter for high-voltage measurements \cite{Spahn2026RamanHV} . Here we demonstrate this scheme for frequency measurements performed on the \Sone, \Pthree, $\mathrm{D}_\mathrm{3/2,5/2}$ off-resonant $\Lambda$-scheme in $^{88}$Sr$^+$, which is similar to the early work of \cite{Kurth.1995} and later by \cite{Yamazaki.2008} $^{40,43}$Ca$^+$ ions in an ion trap. All three possible Raman transitions ($\mathrm{S}_{1/2}\rightarrow\mathrm{D}_\mathrm{J}$ and $\mathrm{D}_{3/2}\rightarrow\mathrm{D}_{5/2}$) are investigated and compared with the results of dipole transition measurements \cite{palmes2025Sr}. The $\mathrm{S}_{1/2}\rightarrow\mathrm{D}_{5/2}$ transition is an ideal benchmark candidate, as it is a well-known clock transition \cite{dube2017absolute}. The $\mathrm{D}_{3/2}\rightarrow\mathrm{D}_{5/2}$ transition was measured employing two consecutive Raman transitions, allowing for a first direct and Doppler-free measurement between two fine-structure states in collinear laser spectroscopy.

\subsection{\label{ssec:CLS}Collinear Laser Spectroscopy} 

In CLS, a beam of ions or neutral atoms is superimposed with a collinear and/or anticollinear laser beam to perform laser spectroscopy. As the ions move with velocity $\upsilon$, the resonance frequency $\nu_\mathrm{c/a}$ measured in the collinear or anticollinear direction in the laboratory frame is Doppler-shifted from the rest-frame transition frequency $\nu_0$ to
\begin{equation}
    \label{eq:dshift_(a)col_v}
    \nu_\mathrm{c/a} = \nu_0 \cdot \gamma (1\pm\beta),
\end{equation}
with the Lorentz factor $\gamma=1/\sqrt{1-\beta^2}$ and $\beta = \upsilon/\mathrm{c}$. $\mathrm{c}$ is the speed of light in vacuum. To reduce Doppler broadening via Doppler compression, the ions are typically electrostatically accelerated to $10-60\,$keV \cite{konig2020new}. By accelerating the ions electrostatically, the energy width is kept constant, reducing the velocity width and thus the Doppler width. For an ion of mass $m$ and charge $q$ accelerated by a voltage $U_\mathrm{acc}$ the Doppler-shifted transition frequency is given by
\begin{eqnarray}
    \label{eq:dshift_(a)col_U}
    \nu_\mathrm{c/a} = \nu_0 \cdot \left[1+\frac{qU_\mathrm{acc}}{m\mathrm{c}^2} \left(1\pm\sqrt{1+2\frac{m\mathrm{c}^2}{qU_\mathrm{acc}}}\right) \right].
\end{eqnarray}
This opens up the possibility of matching the resonance condition by tuning the acceleration voltage (Doppler tuning) instead of the laser frequency, which would limit the laser stability, is only possible over a limited range, and is often connected to high dwell times.
Although possible, in practice, extraction of the rest-frame transition frequency $\nu_0$ from the measured transition frequency $\nu_\mathrm{col/acol}$ and the acceleration voltage employing Eq.\,\eqref{eq:dshift_(a)col_U} is limited in precision by systematic uncertainties in the acceleration voltage \cite{Ullmann2017}. Instead, both $\nu_\mathrm{c}$ and $\nu_\mathrm{a}$ are measured in fast alternation (quasi-simultaneously) \cite{Imgr12CPhysRevLett, PaMuCaPhysRevResearch, Krieger2017Be}. The rest-frame transition frequency is then extracted from the product of the two resonant laboratory-frame laser frequencies, assuming the ion velocity is identical for both measurements, since
\begin{equation}
\label{eq:wacolxwcol}
     \nu_\mathrm{c}\cdot\nu_\mathrm{a}= \nu_0^2 \gamma^2 (1+\beta)(1-\beta) =\nu_0^2.
\end{equation}
In this approach, uncertainties in the ion beam energy cancel out.

\subsection{\label{RamanTrans} Stimulated Raman transitions}

\begin{figure}
    \centering
    \includegraphics[width=.85\linewidth]{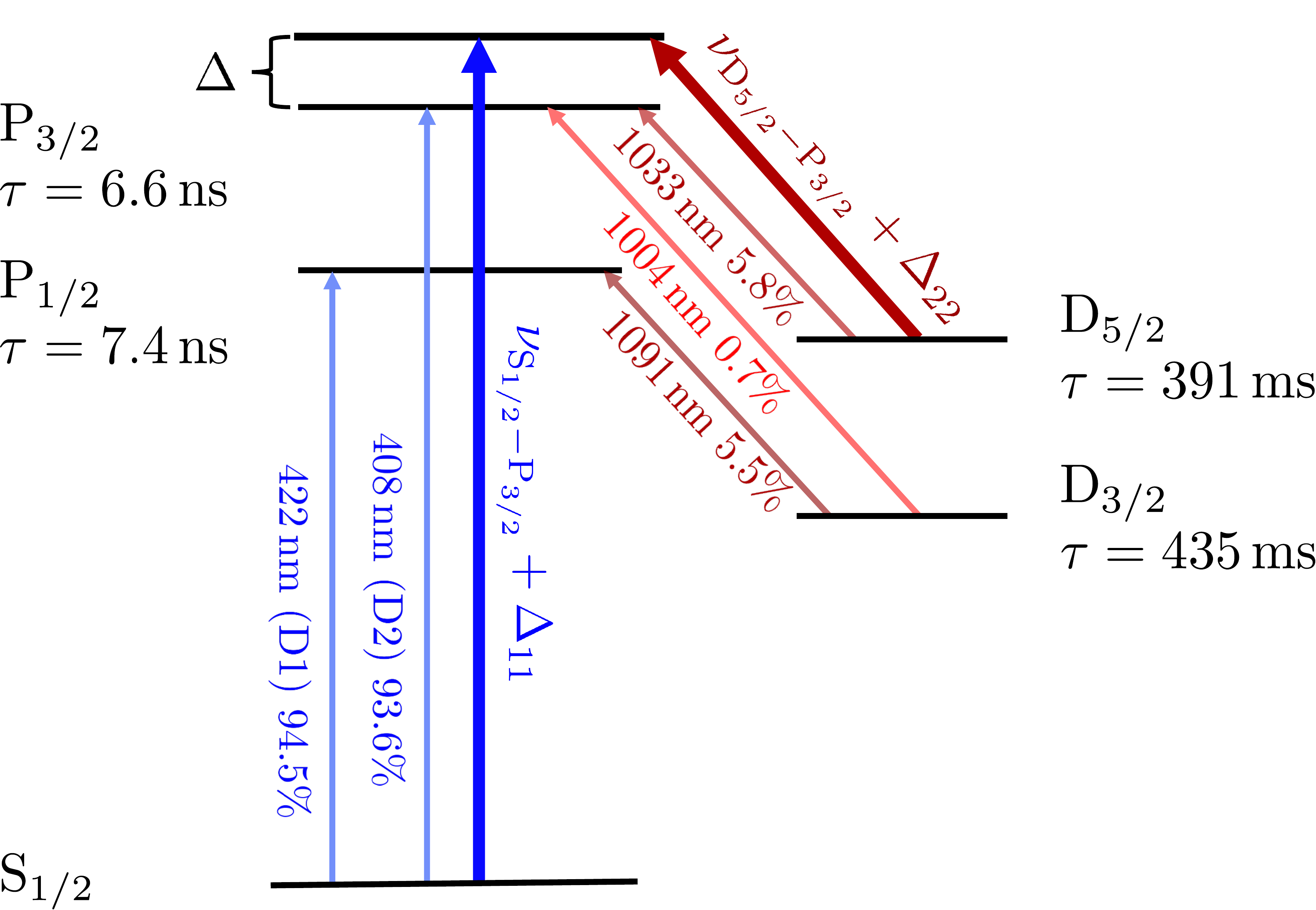}
    \caption{Fine structure level scheme of Sr$^+$. The thin arrows indicate different dipole transitions with their respective wavelength. $\tau$ are the lifetimes of the respective states, and the indicated percentages are the branching ratios for the different decays, taken from \cite{NIST_ASD}.
    The thick arrows indicate a $\mathrm{S}_{1/2}\rightarrow\mathrm{D}_{5/2}$ Raman transition and the dotted line its virtual intermediate state, detuned by $\Delta$.}
    \label{fig:Srlvlscheme_w_RT}
\end{figure}

Stimulated Raman transitions are two-photon transitions from an initial (ground) state to a (metastable) excited state via a detuned virtual intermediate state. A photon from a first laser that is detuned from a dipole-allowed transition from the ground state $\ket{1}$ to a real intermediate state $\ket{3}$ by the single-photon detuning $\Delta_{11}$, is absorbed to excite the atom to a virtual intermediate state. Simultaneously, a stimulated emission of a second photon is induced by a second laser, which is detuned from a dipole-allowed transition from a metastable $\ket{2}$ state to the same real intermediate state by $\Delta_{22}$. This is schematically shown for the $\mathrm{S}_{1/2}\rightarrow\mathrm{D}_{5/2}$ Raman transition in Sr$^+$ in Fig.\,\ref{fig:Srlvlscheme_w_RT}.
If the two-photon resonance condition $\Delta_{11} = \Delta_{22} := \Delta$ is met, this results in a two-photon Rabi oscillation between the ground state and the metastable state with the two-photon Rabi frequency 
\begin{equation}
    \Omega_\mathrm{R} = \frac{\Omega_{11}\Omega_{22}^*}{2\Delta}.
\end{equation}
Here $\Omega_{nj}$ is the Rabi frequency of the dipole transition from the level $\ket{n}$ to the level $\ket{3}$ addressed by the first ($j=1$) or second ($j=2$) laser. The suppression of $\Omega_\mathrm{R}$ with $1/\Delta$, $\Delta \gg \Omega_{ij}$, results in two major differences compared to dipole transitions:
Firstly, the linewidth of a Raman transition is, for realistic laser intensities and detunings, of only a few hundred kilohertz, compared to tens of megahertz for dipole transitions. 
Secondly, the two-photon Rabi oscillation is significantly smaller. Hence, for typical interaction times in CLS of a few microseconds, higher laser powers are required to achieve efficient population transfer. The latter implies that AC-Stark shifts have to be taken into account when measuring transition frequencies. This is due to the linear dependence of the Rabi frequency $\Omega_{nj}$ on the power $P_j$ of laser $j$.

Taking into account the AC-Stark shifts of both lasers, the two-photon resonance condition is given by 
\begin{equation}
    \delta-\delta_\mathrm{AC}=0,
    \label{eq:twophotres_w_AC}
\end{equation}
where $\delta = \Delta_{11} - \Delta_{22}$ is the two-photon detuning and $\delta_\mathrm{AC} = \Omega_1^\mathrm{AC}-\Omega_2^\mathrm{AC}$ is the difference in AC-Stark shifts induced by both lasers in the first dipole transition ($n=1$, $\ket{1}\rightarrow\ket{3}$) and second dipole transition ($n=2$, $\ket{2}\rightarrow\ket{3}$). For a large detuning $\Delta$, the AC-Stark shift is given in first order by
\begin{equation}
    \Omega_n^{AC} = \sum_{j=1,2}\frac{|\Omega_{nj}|^2}{4\Delta_{nj}}, \quad n \in \{1,2\}. 
    \label{eq:Omega_AC}
\end{equation}
If the two lasers address different fine-structure transitions, as is the case for the example shown in Fig.\,\ref{fig:Srlvlscheme_w_RT} and in all transitions investigated in this work, the far off-resonant contributions with $n\neq j$ can be omitted, as $\Delta_{n, j\neq n} \gg \Delta_{n, j=n}$.

Note that while both lasers couple to dipole transitions, the ground-state to metastable state transition is a dipole-forbidden transition since both levels must have the same parity, i.e., this scheme provides access to transitions that are otherwise not accessible via CLS. 

\subsection{\label{ssec:QuasiSim}Quasi-Simultaneous Collinear-Anticollinear Raman Spectroscopy}

In CLS on single-photon transitions, quasi-simultaneous collinear-anticollinear measurements as described in Sec.\,\ref{ssec:CLS} are a well-established method to determine absolute transition frequencies.
In the following, we will describe a scheme that allows expanding this method to collinear Raman spectroscopy. This is achieved by keeping the first laser at a fixed frequency to fix the detuning, and thus the rest-frame resonance frequency of the second laser, and using the second laser in both directions to determine the ion velocity, as it is done for single-photon transitions.

By fixing the frequency $\nu_{\mathrm{col/acol}}^{L1}$ and the direction (collinear or anticollinear) of the first laser, and choosing the collinear and anticollinear laser frequencies $\nu_{\mathrm{col}}^{L2}$ and $\nu_{\mathrm{acol}}^{L2}$ of the second such that the resonances recorded in both directions align in voltage space, the detuning $\Delta$ of the Raman transition, defined by $\nu_{\mathrm{col/acol}}^{L1}$ and the energy of the addressed ions, is fixed. Thus, the two-photon resonance condition for $\nu_{\mathrm{col}}^{L2}$ and $\nu_{\mathrm{acol}}^{L2}$ in the ion rest frame are identical. The resonant rest-frame frequency of the second laser $\nu_{\mathrm{R,}}^{L2}$ can then be obtained by recoding the resonance in both directions in fast alternation via Doppler tuning and fitting the resonance positions $U_{\mathrm{col/acol}}$ in voltage space.
\begin{equation}
    \nu_{\mathrm{R,}}^{L2} = \sqrt{\nu_{\mathrm{acol}}^{L2}\cdot\left(\nu_{\mathrm{col}}^{L2} + \Delta D\cdot\Delta U + \delta \nu_{\mathrm{AC,}}^{L2}\right)}\mathrm{.}
\end{equation}
Compared to Eq.\,\eqref{eq:wacolxwcol}, the term $\Delta D\cdot \Delta U$ is the first-order correction for the misalignment $\Delta U = U_{\mathrm{acol}} - U_{\mathrm{col}}$. $\Delta D$ is the difference in differential Doppler factors 
\begin{equation}
    \begin{aligned}
        \Delta D &= D(\nu_{\mathrm{acol}}^{L1}) -  D(\nu_{\mathrm{col}}^{L2})\\
        &= \frac{\partial \nu_{\mathrm{acol}}^{L1}}{\partial U} - \frac{\partial \nu_{\mathrm{col}}^{L2}}{\partial U}\mathrm{.}
    \end{aligned}
\end{equation}
The second correction term $\delta \nu_{\mathrm{AC}}^{L2}=\delta \nu_\mathrm{AC,\,acol}^{L2} - \delta \nu_\mathrm{AC,\,col}^{L2}$ corrects for differences in the AC-Stark shifts $\delta \nu_\mathrm{AC,\,col/acol}^{L2}$ of the second laser used in collinear-anticollinear geometry, which can be caused by changes in the laser power or the laser beam diameter. $\delta \nu_\mathrm{AC,\,col/acol}^{Li}$ are the measured AC-Stark shifts, which include the negative sign for the shift induced in the second dipole transition in the definition of $\delta_\mathrm{AC} = \Omega_1^\mathrm{AC}-\Omega_2^\mathrm{AC}$ in Eq.\,\eqref{eq:twophotres_w_AC}.\\

The relative Doppler shift $\gamma(1+\beta)$, equivalent to the ion velocity, can then be extracted using Eq.\,\ref{eq:dshift_(a)col_v} under the assumption that all lasers interact with ions of the same energy and hence velocity: 
\begin{equation}
    \gamma(1+\beta) = \nu_{\mathrm{R}}^{L2} / \nu_{\mathrm{acol,}}^{L2}.
\end{equation}
This can then be used to calculate the rest-frame frequency of the first laser $\nu_{\mathrm{R}}^{L1}$ from the (anti)collinear laboratory frequency $\nu_{\mathrm{col/acol}}^{L1}$, 
\begin{equation}
   \nu_{\mathrm{R}}^{L1} = \nu_{\mathrm{col/acol}}^{L1} \cdot \left(\nu_{\mathrm{R}}^{L2} / \nu_{\mathrm{acol}}^{L2}\right)^{\mp1} \mathrm{.}
\end{equation} 
Finally, the rest-frame transition frequency from the ground state to the metastable, given by the difference of the two rest-frame laser frequencies, corrected for the AC-Stark shifts, can be calculated:
\begin{equation}
    \begin{aligned}
        \nu_{0}^{\ket{1}\rightarrow\ket{2}} & =  \nu_{\mathrm{R}}^{L1} - \nu_{\mathrm{R}}^{L2} \\
        &- (\delta \nu_\mathrm{AC,\,acol}^{L1} + \delta \nu_\mathrm{AC,\,acol}^{L2}).
    \end{aligned}
    \label{eq:RamanACCAf0}
\end{equation}

\section{\label{sec:ExpSet}Experimental Setup}
The Collinear Apparatus for Laser Spectroscopy and Applied Science (COALA) at the Institute for Nuclear Physics at TU Darmstadt is designed for high-precision CLS measurements on stable isotopes and to develop novel experimental techniques for online experiments \cite{konig2020new}.

\begin{figure*}
    \centering
    \includegraphics[width=1\linewidth]{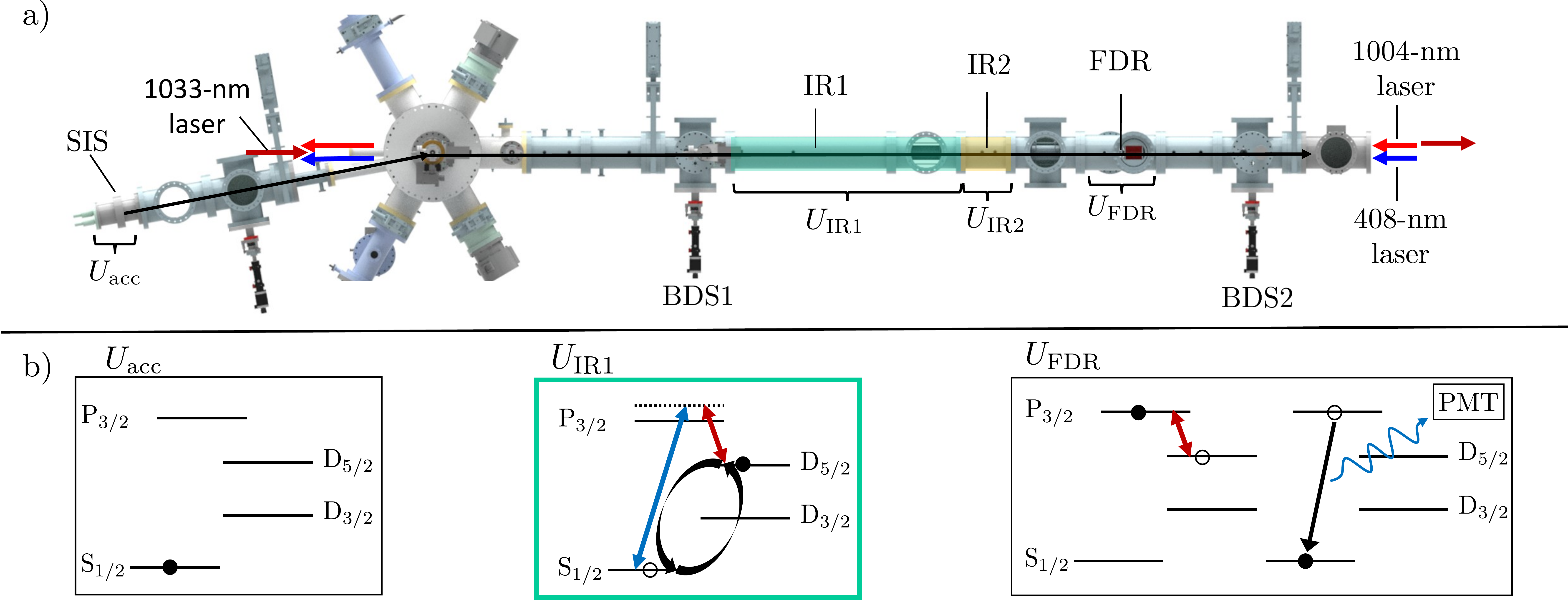}
    \caption{
    a) Schematics of the COALA beamline. The ions are produced in the surface ionization source (SIS) placed on a high voltage $U_\mathrm{acc}$, accelerated to ground potential, and superimposed with the laser beams using electrostatic ion optics and the two Beam diagnostic stations (BDS1, BDS2). Two interaction regions, IR1 and IR2, are used to drive Raman transitions via Doppler tuning by floating the interaction regions to $U_\mathrm{IR1}$ and $U_\mathrm{IR2}$, respectively. The fluorescence detection region (FDR) is equipped with photomultiplier tubes, used to detect the 408-nm photons from the $\mathrm{P}_{3/2}\rightarrow\mathrm{S}_{1/2}$ decay.
    b) Measurement scheme for the $\mathrm{S}_{1/2}\rightarrow\mathrm{D}_{3/2}$ Raman transition. Ions leaving the ion source primarily populate the ground state. They are then transferred to the $\mathrm{D}_{3/2}$ state by matching the two-photon resonance condition of the Raman transition via Doppler tuning of  $U_\mathrm{IR1}$.
    The FDR is floated to a fixed voltage $U_\mathrm{FDR}$ that matches the $\mathrm{D}_{3/2}\rightarrow\mathrm{P}_{3/2}$ dipole condition. Ions excited into the $\mathrm{D}_{5/2}$ state via the Raman transition are resonantly excited into the $\mathrm{P}_{3/2}$ state, enabling spectroscopy on the Raman transition by detecting the 408-nm photons emitted from the $\mathrm{P}_{3/2}\rightarrow\mathrm{S}_{1/2}$ decay.
    }
    \label{fig:Beamline+MScheme}
\end{figure*}

Laser spectroscopy on $\mathrm{Sr}^+$ ions was performed using three continuous-wave lasers: A first Ti:Sapphire (Ti:Sa) laser (Sirah Matisse-2), pumped by a frequency-doubled, diode-pumped $\mathrm{Nd:YVO}_4$ laser (Spectra-Physics Millennia eV), is used to produce $1004$-nm or $1033$-nm light. A second identical Ti:Sa laser is used in combination with a frequency doubling stage (Spectra-Physics Wavetrain) equipped with an LBO crystal to produce 408-nm light. The 408-nm beam is modulated using an AOM, which induces a fixed frequency shift of $200\,\mathrm{MHz}$.
Both Ti:Sa lasers are short-term stabilized to a reference cavity. The reference cavities are again long-term stabilized to a frequency comb (Menlo Systems FC1500-250-WG), enabling frequency measurements at sub-$50\,$kHz accuracy in combination with a wavemeter (HighFinesse WS8-2, stabilized to a He:Ne laser). This accuracy is limited by the laser linewidth. 
Additionally, a tunable diode laser at $1033\,$nm (Toptica DL Pro) is used. Its frequency is stabilized to a second wavemeter (High Finesse WSU 30, calibrated to the same He:Ne laser as the WS8-2) via a PID loop and monitored with the WS8-2. The two beams from the Ti:Sa lasers are superimposed using a long-pass dichroic mirror and guided through the beamline in anticollinear geometry, where the diode laser beam is copropagating with the ion beam.

The COALA beamline is shown in Fig.\,\ref{fig:Beamline+MScheme}(a): $\mathrm{Sr}^+$ ions are produced in a surface ionization source, a resistively heated graphite crucible filled with Sr and placed on a $U_\mathrm{acc}=20\,$kV high-voltage platform. The $20\,$kV acceleration voltage is generated with a Heinzinger PNChp 20000-10 high-voltage power supply, monitored using a high-voltage divider and a Keysight 3458A multimeter, and long-term stabilized using the multimeter readout and a custom DAC \cite{Koenig2024HV}.
From this high-voltage platform, the ions are accelerated to ground potential and bent by $10^{\circ}$ into the main beamline and aligned with the laser beams using electrostatic steering electrodes. A quadrupole doublet is used to optimize the ion beam profile, which is monitored alongside the beam alignment using two beam diagnostic stations. They are equipped with an adjustable iris aperture as well as a Faraday cup and MCP connected to a phosphorus screen that can be lowered into the ion beam. The phosphorus screen faces a mirror that reflects the image of the phosphorus screen into a CCD camera. 
The Raman transition is driven in a $1.2\,$m long interaction region IR1 in the center of the beamline, consisting of a metallic tube that can be floated to a separate voltage $U_\mathrm{IR1}$, allowing for a well-defined interaction potential. An optional second Raman transition can be driven in a second, $0.3$-m long interaction region IR2 (interaction potential $U_\mathrm{IR2}$), located immediately downstream of IR1. 
Both interaction regions are shielded against magnetic fields using two layers of mu-metal foils, reducing the Earth's magnetic field by about one order of magnitude. The voltages $U_\mathrm{IR1}$ and $U_\mathrm{IR2}$ are applied by amplifying a 0--10\,V voltage of a DAC unit by a factor of $50$  using a Kepco BOP500M. 
The COALA fluorescence detection chamber (FDR) is finally used for the optical detection of the fluorescence photons originating from the $\mathrm{P}_{3/2}\rightarrow\mathrm{S}_{1/2}$ decay, using one lens-based and one mirror-based segment \cite{Muller.2025,Koenig2024HV}, which focus the fluorescence light into photomultiplier tubes (PMT)(Sens-Tec P25PC with UV glass; 25\% quantum efficiency at 420\,nm).

\section{\label{sec:ScanProcedure}Scanning procedure for Raman Transitions}

The resonances of the Raman transitions were recorded by driving the Raman transition in IR1 or IR2 and
consecutively probing the population of the targeted metastable state $\mathrm{D}_{3/2,5/2}$ in the FDR, as illustrated in Fig.\,\ref{fig:Beamline+MScheme}(b) for the $\mathrm{S}_{1/2}\rightarrow\mathrm{D}_{5/2}$ transition. 
If the resonance condition of the Raman transition is met in IR1, ions initially in the ground state are transferred to the metastable $\mathrm{D}_{J}$ (Fig.\,\ref{fig:Beamline+MScheme}(b), center panel), and the population of the probed state increases. The metastable state is probed by resonantly driving the $\mathrm{D}_{J}\rightarrow\mathrm{P}_{3/2}$ dipole transition (Fig.\,\ref{fig:Beamline+MScheme}(b), right panel).
Ions in the $\mathrm{D}_{J}$ state are excited into the $\mathrm{P}_{3/2}$ state ($\tau=6.6\,$ns) in the FDR and decay into the $\mathrm{S}_{1/2}$ ground state, emitting fluorescence photons detected by the PMTs, resulting in an increase in the PMT counts.

Scanning the Raman transition and resonantly probing the metastable state requires different rest-frame laser frequencies of the 1004/1033-nm laser, as illustrated by the different lengths of the red arrows indicating the 1033-nm laser in Fig.\,\ref{fig:Beamline+MScheme}(b). Hence, either a third laser is required, or, as demonstrated here, Doppler tuning can be employed.
By applying different voltages $U_\mathrm{IR1/IR2}$ to IR1/IR2 and $U_\mathrm{FDR}$ to the FDR, while keeping the laser frequencies fixed, the ion velocity, and thus the Doppler-shifted rest-frame laser frequencies are modified through the interaction potential. This enables simultaneous driving of the Raman transition and resonant probing of metastable state via the $\mathrm{D}_{J}\rightarrow\mathrm{P}_{3/2}$ transition with a single laser.
The resonance is then recorded by scanning $U_\mathrm{IR1/IR2}$. Since the two lasers have different frequencies or directions, scanning $U_\mathrm{IR1/IR2}$ scans the difference of the two laser frequencies in the ion rest frame and thus the two-photon resonance condition. Recording the PMT counts as a function of the scanned voltage therefore results in a peak in the recorded signal when the resonance condition of the Raman transition is met.

Since the PMT detects UV light and high laser powers are required to drive the Raman transition, which results in a high laser-induced background, the 408-nm beam is switched off using the AOM, while the ions are probed in the FDR. Doing so eliminates the corresponding background and increases the signal-to-noise ratio (SNR) by a factor of 6.\\

Spectroscopy on the $\mathrm{D}_{5/2}\rightarrow\mathrm{D}_{3/2}$ Raman transition requires a prior population of the $\mathrm{D}_{5/2}$ state. This was achieved by first driving the $\mathrm{S}_{1/2}\rightarrow\mathrm{D}_{5/2}$ Raman transition at a fixed interaction potential $U_\mathrm{IR1}$ in IR1 and consecutively measuring the $\mathrm{D}_{5/2}\rightarrow\mathrm{D}_{3/2}$ Raman transition in IR2 by scanning $U_\mathrm{IR2}$. 
Compared to optical pumping through the $\mathrm{P}_{3/2}$ state, this approach does not simultaneously populate the $\mathrm{D}_{3/2}$ state, which would reduce the contrast in population when detecting the $\mathrm{D}_{5/2}\rightarrow\mathrm{D}_{3/2}$ Raman transition via resonant probing of the $\mathrm{D}_{3/2}$ state in FDR.
Furthermore, since in IR1 only ions of an energy width corresponding to the width of the first Raman transition are selected, this measurement is in principle Doppler-free.

\section{\label{ssec:LineShape}Resonance Line shape}

\begin{figure*}[!t]
     \centering
    \includegraphics[width=1\linewidth]{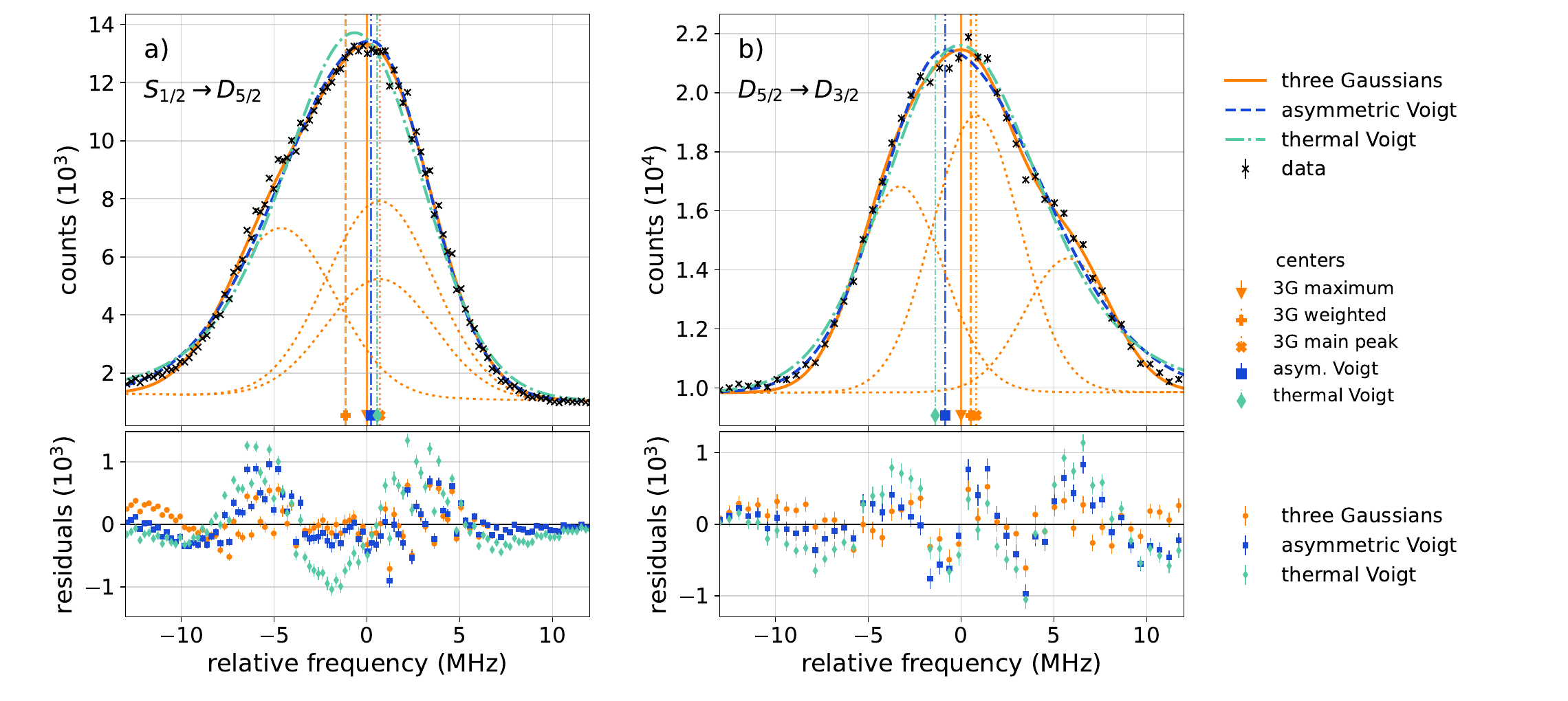}
    \caption{Comparison of different fitted line shapes  for (a) the $\mathrm{S}_{1/2}\rightarrow\mathrm{D}_{5/2}$  and (b) the $\mathrm{D}_{5/2}\rightarrow\mathrm{D}_{3/2}$ Raman transition. The black data points and error bars indicate the PMT counts and their statistical uncertainties. The colored lines indicate the fit results and the dotted lines the three peaks of the triple Gaussian line shape. The vertical lines in respective colors indicate the different peak centers, with different methods being used for the triple Gaussian line shape. At the bottom, the residuals of the different fits are shown. The frequencies on the $x$-axis are given relative to $436\,373\,693\,$MHz in (a) and $8\,405\,350\,$MHz in (b).}
    \label{fig:Plot_Lineshapes}
\end{figure*}

Figure\,\ref{fig:Plot_Lineshapes} shows example spectra of the measured $\mathrm{S}_{1/2} \rightarrow\mathrm{D}_{5/2}$ and $\mathrm{D}_{5/2}\rightarrow\mathrm{D}_{3/2}$ Raman transitions. The $\mathrm{S}_{1/2}\rightarrow\mathrm{D}_{5/2}$ resonance was measured using lasers anti-parallel to the ion beam at $1033\,$nm and $408\,$nm, and the $\mathrm{D}_{5/2}\rightarrow\mathrm{D}_{3/2}$ resonance was driven by the co-propagating diode laser at $1033\,$nm and the counter-propagating Ti:Sa laser at $1004\,$nm.
While the intrinsic shape of a Raman transition is expected to be a Lorentzian with a width of a few hundred kHz \cite{neumann2020raman}, the observed line shape distinctly differs.

Since spectroscopy on the $\mathrm{S}_{1/2}\rightarrow\mathrm{D}_{3/2}$ and $\mathrm{S}_{1/2}\rightarrow\mathrm{D}_{5/2}$ transitions is performed directly on the ground-state population of the ions leaving the ion source, the line shape is dominated by Doppler broadening. Indeed, fitting a Voigt profile shows a vanishing Lorentzian contribution to the line shape and a Gaussian width of $\sigma = 4.6(1)\,$MHz or $0.61(1)\,$eV. 
A significant asymmetric substructure with indications of a triple-peak structure can be seen in the resonance signal. 
This substructure does not vary with laser intensities, and its width in voltage space is independent of the direction of the 1033-nm laser, indicating that the substructure originates from the ion energy distribution. It is ascribed to different hot spots along the voltage gradient of the heated graphite crucible, resulting in different acceleration potentials of the ions.
Different line shapes were fitted, including a thermal fit (convolution of a Lorentzian with an electrostatically accelerated Maxwell-Boltzmann distribution \cite{Kretzschmar.2004,muller2025qspec}, an asymmetric Voigt profile, where the asymmetry is defined by the ratio of the left- and right-flank widths, as well as a triple-peak structure, consisting of the sum of three different Gaussian peaks with identical widths. All line shapes reproduce the observed peak. The asymmetric and triple-peak line shapes result in similar residuals and reduced chi-square values, while the thermal fit yields a factor of two to four higher reduced chi-square values. 
Since the fit results vary depending on the line shapes and, for the triple-peak line shape, the definition of the line center, a systematic uncertainty was introduced based on the differences between the analysis results using different line shapes (see Sec. \ref{ssec:Lineshape}).


Spectroscopy on the $\mathrm{D}_{5/2}\rightarrow\mathrm{D}_{3/2}$ Raman transition was performed as described in Sec. \ref{sec:ScanProcedure}.
However, the observed resonance has a width of $3.3(1)\,$MHz or $0.32(1)\,$eV and exhibits a triple-peak structure. The same line shapes as for the  $\mathrm{S}_{1/2}\rightarrow\mathrm{D}_{J}$ transitions were fitted, and the triple-peak line shape was found to best reproduce the measured data. 
Comparing the fitted triple-peak to the one fitted for the $\mathrm{S}_{1/2}\rightarrow\mathrm{D}_{J}$ transitions shows significant differences in the relative intensities and positions of the sub-peaks. This indicates that the substructure in the $\mathrm{D}_{5/2}\rightarrow\mathrm{D}_{3/2}$ resonance originates from a different effect than the substructure in the $\mathrm{S}_{1/2}\rightarrow\mathrm{D}_{J}$ resonances. Measurements of the Lamb-dip in the ground state population via collinear Raman saturation spectroscopy using the same setup show that neither the laser linewidth, AC-Stark broadening, time-of-flight broadening, nor variations in the interaction potential explain a broadening beyond $2\,$MHz \cite{Spahn2026RamanHV, spahn2026pioneeringRaman}.
A small detuning of $175\,$MHz from the $\mathrm{D}_{J}\rightarrow\mathrm{P}_{3/2}$ dipole transitions had to be chosen for the $\mathrm{D}_{5/2}\rightarrow\mathrm{D}_{3/2}$ transition to achieve a significant population transfer due to the short interaction time in IR2, as well as the $\mathrm{D}_{J}\rightarrow\mathrm{P}_{3/2}$ dipole transitions, and thus, the $\mathrm{D}_{5/2}\rightarrow\mathrm{D}_{3/2}$ Raman transition being slower.
Simulations including the finite lifetime of the metastable $\mathrm{D}_\mathrm{J}$ states using \textit{qspec} \cite{muller2025qspec} indicate that choosing a too-small detuning can result in line shapes with asymmetric side peaks, which are artifacts of the two-photon \textit{sinc} Rabi-oscillation in frequency space. This might be the case at the chosen parameters. Furthermore, if the IR1 potential is not fixed at the center of the $\mathrm{S}_{1/2}\rightarrow\mathrm{D}_{5/2}$ resonance, a different ion energy than the center of the ion energy distribution is selected and probed in the $\mathrm{D}_{5/2}\rightarrow\mathrm{D}_{3/2}$ interaction. If so, the probed ions have an energy different from the one obtained via the optical voltage calibration. Although this was verified by recording the resonance in the first interaction region between the measurement sets, small deviations cannot be excluded. To account for these effects, an additional uncertainty of $0.5\,$MHz is introduced, based on results obtained using different fit routines and at different interaction potentials $U_\mathrm{IR1}$.

\section{\label{Results}Results}

\subsection{AC-Stark shifts}
Transition frequency measurements were performed at laser powers of up to $3\,$mW, $300\,$mW, and $20\,$mW for the frequency-doubled Ti:Sa laser ($408\,$nm), the Matisse used at $1004\,$nm/$1033\,$nm, and the 1033-nm diode laser, respectively.
Hence, AC-Stark shifts are expected to be in the kHz to low MHz range and can no longer be omitted.

\begin{figure}
    \centering
    \includegraphics[width=0.95\linewidth]{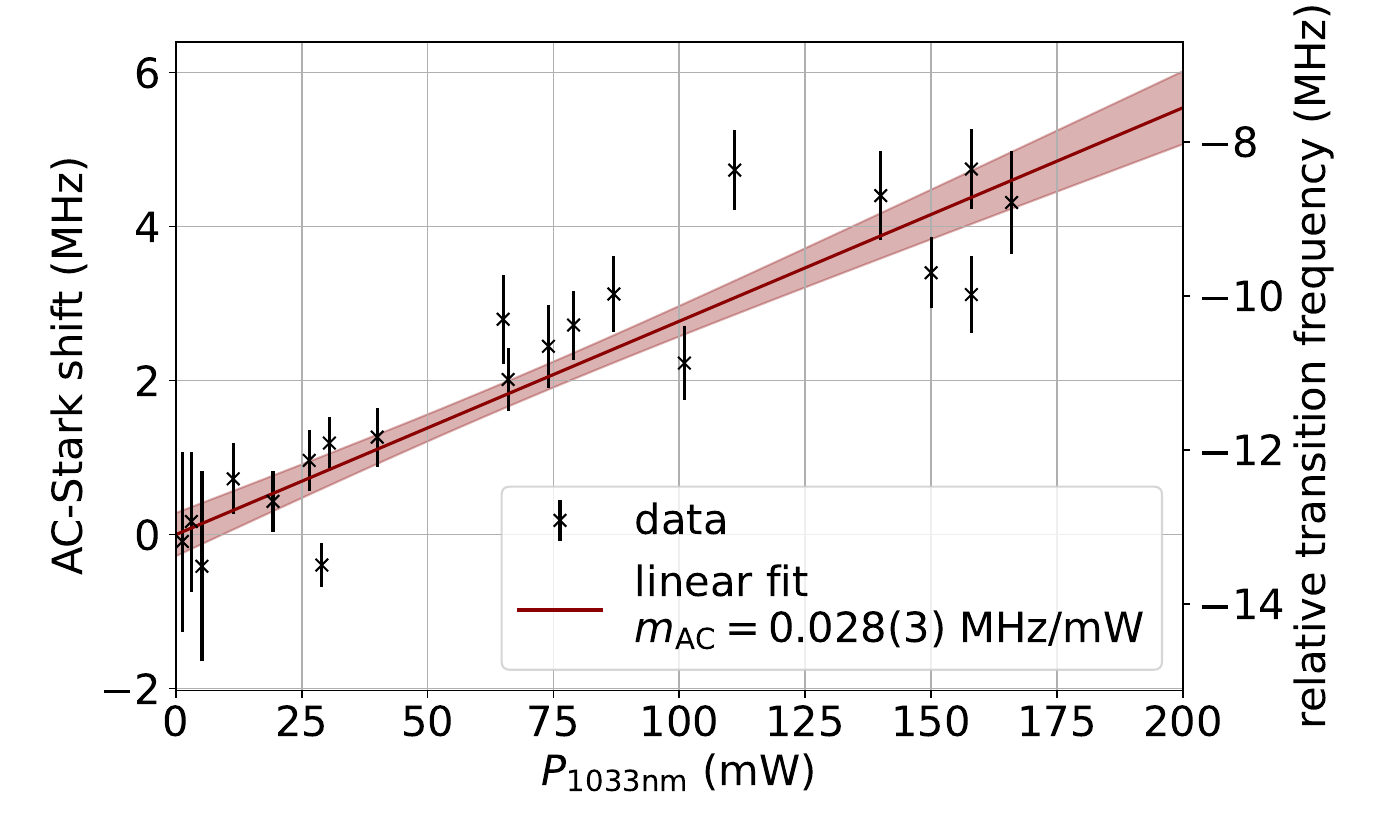}
    \caption{AC-Stark shift and relative transition frequency of the $\mathrm{S}_{1/2}\rightarrow\mathrm{D}_{5/2}$ transition depending on the power $P_{1033}$ of the 1033-nm laser. The measurements were performed in IR1 at a detuning of \SI{-483(1)}{MHz} and at a  fixed power of the 408-nm laser of \SI{155(10)}{\micro W}. The transition frequencies on the right y-axis are given relative to $444\,779\,044\,$\SI{}{MHz}. The black error bars indicate the experimental data and the fit uncertainties of the transition frequencies. To investigate the AC-Stark shift, a linear function was fitted. The fit result is shown with its \textcolor{red}{1\,$\sigma$} confidence interval in red. The slope of $m_\mathrm{AC}=0.028(3)\,$MHz/mW corresponds to the power-dependent AC-Stark shift, and the offset of $-13.1(3)\,$MHz at  $P_{1033}=0$ in the relative transition frequencies corresponds to the offset of the wavemeter used for these measurements.}
    \label{fig:AcStark_1033}
\end{figure}

To correct the measured transition frequencies, measurements on the $\mathrm{S}_{1/2}\rightarrow\mathrm{D}_{3/2}$ and $\mathrm{S}_{1/2}\rightarrow\mathrm{D}_{5/2}$ Raman transitions in $^{88}$Sr$^+$ were performed at different laser powers. For the 1004-nm and 1033-nm transitions, the Matisse was used to cover a wider range of laser powers. As an example, the dependence of the $\mathrm{S}_{1/2}\rightarrow\mathrm{D}_{5/2}$ transition frequency at a detuning of $-483(1)\,$MHz on the 1033-nm laser power is shown in Fig.\,\ref{fig:AcStark_1033}.

As expected, a linear dependence can be observed, and a linear function was fitted to the experimental data. The results of those fits for the respective dipole transitions are summarized in Tab.\,\ref{tab:resACStark}, including the beam diameters and detunings used in the experiment. The beam diameters correspond to $4\sigma$ of a Gaussian fit to the beam intensity, measured using a Thorlabs BC106-VIS beam profiler. Measurements of the laser beam diameters were performed at different positions along the beam path and extrapolated for the two interaction regions.

\begin{table}
    \centering
       \caption{Results of the performed AC-Stark measurements: $m_\mathrm{AC}$ is the slope and y-intercept $b$ of the linear fit to the measured transition frequencies, $\Delta_0$ and $d_0$ are the laser detuning and the beam diameter during the measurements.}
    \begin{tabular}{c c c c}
        \hline\hline
        transition \quad \quad & $m_\mathrm{AC}$ (MHz/mW)  \quad \quad & $\Delta_0$ (MHz)  \quad \quad & $d_0$ (mm)\\\hline
        $\mathrm{D}_{5/2}\rightarrow\mathrm{P}_{3/2}$& $0.028(3)$ & $-483(1)$ & $3.3(3)$ \\
        $\mathrm{D}_{3/2}\rightarrow\mathrm{P}_{3/2}$& $-0.0031(1)$ & $1091(1)$ & $2.7(1)$ \\
        $\mathrm{S}_{1/2}\rightarrow\mathrm{P}_{3/2}$& $-0.41(13)$ & $-298(1)$ & $1.8(1)$ \\
        \hline\hline
    \end{tabular}
    \label{tab:resACStark}
\end{table}

No frequency comb was used for these measurements, as only the intensity-dependent relative frequency shift was extracted. Hence, the $y$-intercepts of the fits are shifted due to the systematic offset of the wavemeter. Alternatively, one could perform these measurements with a frequency comb and then extrapolate for a laser power of zero from the $y$-intercept.

The measured AC-Stark shift is averaged over all ions interacting with the two laser beams driving the Raman transition. It thus depends highly on the spatial intensity distributions of both the laser- and ion beams and their overlap. Hence, a direct comparison of the measured slopes to theoretical predictions is not possible. However, by normalizing the slopes to $\Delta_0$ and $d_0$, the ratios of the normalized slopes, which are proportional to the square of the dipole moments $|d_{\mathrm{eff}, n}|$ of the respective dipole transitions, can be compared to the theoretical dipole moments. This results from Eq.\,\eqref{eq:Omega_AC} and 
\begin{equation}
    |\Omega_{nj}| = \frac{|d_{\mathrm{eff}, n}|}{\hbar}\sqrt{\frac{4P_j}{\pi\epsilon_0 \mathrm{c} w_j^2}},
\end{equation}
where $d_{\mathrm{eff}, n}$ is the effective dipole moment of the transition $n$.
The latter can be obtained from the transition frequencies and lifetimes of the transitions \cite{dunning2015coherent}, which were taken from \cite{NIST_ASD}. 
For the $\mathrm{S}_{1/2}\rightarrow\mathrm{P}_{3/2}$ ($408\,$nm) and $\mathrm{D}_{5/2}\rightarrow\mathrm{P}_{3/2}$ ($1033\,$nm) transitions, the calculated dipole ratio is $3.0$, and for the $\mathrm{D}_{3/2\rightarrow}\mathrm{P}_{3/2}$ ($1004\,$nm) and the 1033-nm $\mathrm{D}_{5/2}\rightarrow\mathrm{P}_{3/2}$ transitions it is $0.16$. The measured ratios are $2.8(9)$ and $0.20(3)$, respectively, yielding good agreement.
The sign of the slope behaves as expected, i.e., it changes when the sign of the detuning changes, and at a given detuning, the sign is opposite for the two involved dipole transitions.

The measured absolute transition frequencies are corrected by scaling the measured AC-Stark shifts with the laser beam diameters $d$ and the detuning $\Delta$, as well as the laser power $P$, at which the measurements were performed:
\begin{equation}
    \delta \nu_\mathrm{AC, 1} = m_\mathrm{AC}\cdot\frac{\Delta_0}{\Delta}\frac{d_0^2}{d^2}\cdot P.
\end{equation}

The expression given in Eq.\,\eqref{eq:Omega_AC} is a first-order approximation of the full AC-Stark shift for $\Omega_{ii} \ll \Delta$, which can be obtained using the dressed-state approach \cite{dunning2015coherent}.
Measurements of the $\mathrm{D}_{5/2}\rightarrow\mathrm{D}_{3/2}$ were conducted at a detuning of $-175\,$MHz. The approximation $\Omega_{ii} \ll \Delta$ no longer holds and the second-order AC-Stark shift must be taken into account. A Taylor series expansion up to the second order for  $\Omega_{ij} \ll \Delta$, and neglecting shifts induced by the far-off-resonant laser, yields 
\begin{equation}
    \label{eq:Omega_AC_full}
    \Omega_n^{AC} = \frac{|\Omega_{nn}|^2}{4\Delta} - \frac{|\Omega_{nn}|^4}{16\Delta^3}. 
\end{equation}
The correction of the AC-Stark shift up to the second order is hence given by $\delta \nu_\mathrm{AC} = \delta \nu_\mathrm{AC,1}\left(1-\big|\frac{\delta \nu_\mathrm{AC,1}}{\Delta}\big| \right)$, 
where $\delta \nu_\mathrm{AC,1} = \frac{|\Omega_{ii}|^2}{4\Delta}$ is the first-order correction.
For the $\mathrm{D}_{5/2}\rightarrow\mathrm{D}_{3/2}$ transition measurements, the second-order contribution is several $100\,$kHz. For all other transitions, the second-order contribution was included in the analysis as well but is negligible compared to the other systematic uncertainties discussed below. The third-order corrections were calculated to be a few Hz and were not included.

\subsection{\label{secTransitionfreqs}Transition frequency measurement}

\subsubsection{\label{ssec:QuasiSim_freqs}Via quasi-simultaneous collinear-anticollinear measurements}
As it is a well-known clock transition, the $\mathrm{S}_{1/2}\rightarrow\mathrm{D}_{5/2}$ transition is an ideal candidate to benchmark collinear Raman spectroscopy. To test the quasi-simultaneous collinear-anticollinear method, the $\mathrm{S}_{1/2}\rightarrow\mathrm{D}_{5/2}$ transition was investigated at a detuning of $\-680\,$MHz in this scheme by using the diode laser at $1033\,$nm in the collinear direction and the Matisse laser at $1033\,$nm  in the anticollinear direction. The anticollinear 408-nm laser was kept at a fixed frequency to fix the detuning in both directions; see Sec.\,\ref{ssec:QuasiSim}. The 1033-nm laser frequencies were adjusted so that the collinear and anticollinear resonances align in voltage space. The resonance positions $U_{\mathrm{col}}$, $U_{\mathrm{acol}}$ were determined via Doppler tuning (see Sec.\,\ref{sec:ExpSet}). 
Anticollinear-collinear and collinear-anticollinear pairs were measured in fast alternation with a measurement time of approximately two minutes per spectrum. 
Typical laser powers of $P_{408}$=2-3\,mW, $P_{1033}$=10-15\,mW in collinear direction, and $P_{1033}$=20-45\,mW in anticollinear direction were used. The laser powers of the 1033-nm lasers were chosen such that the signal intensities match in both directions.
The extracted and AC-Stark shift-corrected rest-frame transition frequencies are consistent for different days and laser powers. The obtained weighted average is $444\,779\,044.21\,$MHz with a standard deviation of $0.64\,$MHz and a statistical uncertainty of $80\,$kHz.
Including the systematic uncertainties listed in Tab.\,\ref{tab:FreqResults} and discussed below, this results in a transition frequency of 
\begin{equation*}
    \nu_{0, \mathrm{S}_{1/2}\rightarrow\mathrm{D}_{5/2}} = 444\,779\,044.21(8)_\mathrm{stat}(34)_\mathrm{sys}\,\mathrm{MHz.}
\end{equation*}
This is in excellent agreement with the value of $444\,779\,044.095\,485\,27\,(75)\,$\SI{}{MHz} measured by Dubé \textit{et al.} using an atomic clock \cite{dube2017absolute}. The precision is limited by uncertainties in the AC-Stark shift and in the frequency of the diode laser. A reduction of the systematic contributions could be achieved by measuring the AC-Stark shift with a frequency comb, which requires a modification of the frequency-comb setup to allow for simultaneous measurements of three laser frequencies.

\subsubsection{\label{ssec:OpticalVoltCal_freqs}Via Optical Voltage Calibration}

Quasi-simultaneous collinear-anticollinear transition frequency measurements of the $\mathrm{S}_{1/2}\rightarrow\mathrm{D}_{3/2}$ and the $\mathrm{D}_{5/2}\rightarrow\mathrm{D}_{3/2}$ Raman transitions could not be performed with the available laser setup. As relying on electronic measurements of the applied acceleration voltages $U_\mathrm{acc}$ and $U_{\mathrm{IR1,2}}$ does not account for space charge effects in the ion source, contact potentials between different parts of the beamline, and field penetrations into the interaction regions, a more sophisticated approach is required to determine the ion energy. Therefore, the 408-nm $\mathrm{S}_{1/2}\rightarrow\mathrm{P}_{3/2}$ transition (D2-line) of $^{88}$Sr$^+$ was chosen as a reference to determine the ion energy. Since this rest-frame transition frequency is known at high precision \cite{palmes2025Sr}, the frequency of the $408\,$nm-laser can be used to extract the effective acceleration potential by measuring the D2 transition in anticollinear geometry in the laboratory frame and applying Eq.\,\eqref{eq:dshift_(a)col_U}. 
That value is compared to the electronically measured high voltage to which the ion source is floated and the scan voltage applied to the interaction region. The difference between those values is used to correct the electronically measured acceleration voltage of the subsequent measurement on a Raman transition. 
The rest-frame laser frequencies are calculated using that voltage and Eq.\,\eqref{eq:dshift_(a)col_U}, and the transition frequency is obtained from the difference of the rest-frame laser frequencies, corrected for the AC-Stark shift, analogously to Eq.\,\eqref{eq:RamanACCAf0}.

Measurements on the $\mathrm{S}_{1/2}\rightarrow\mathrm{D}_{5/2}$, $\mathrm{S}_{1/2}\rightarrow\mathrm{D}_{3/2}$, and $\mathrm{D}_{5/2}\rightarrow\mathrm{D}_{3/2}$ transitions were performed at detunings of $665\,$MHz, $-1020\,$MHz, and $175\,$MHz, respectively. 

The obtained transition frequencies and their uncertainties are summarized in Tab.\,\ref{tab:FreqResults}.  
The value for the $\mathrm{S}_{1/2}\rightarrow\mathrm{D}_{5/2}$ transition is in very good agreement with the result of the quasi-simultaneous collinear-anticollinear determination, but the precision of the latter is not matched. 

\sisetup{input-symbols = {()},  
         group-digits  = true} 
\begin{table*}[]
    \centering
    \caption{List of absolute transition frequencies and their different systematic uncertainties for the measured transition frequencies using the quasi-simultaneous collinear-anticollinear method and the optical voltage calibration. Uncertainties below $50\,$kHz are marked as $< 0.1\,$MHz for better comparison. The obtained frequencies are compared to the results reported in \cite{palmes2025Sr}, which were obtained by taking frequency differences of the respective E1 transitions.}
    \begin{tabular}{c S[table-format=10.7] c c c c c c c}
        \hline\hline
        transition & {frequency} & total uncertainty  & stat. & AC-Stark & energy & laser & alignment & line shape \\ 
                   & {(MHz)} & (MHz) & (MHz) & (MHz) & (MHz) & (MHz) & (MHz) & (MHz) \\
        \hline\hline
         $\mathrm{S}_{1/2}\rightarrow\mathrm{D}_{5/2}$ & & & & & & & & \\    
         quasi-simultaneous col.-acol. 
         & 444779044.2 & $0.4$ & $0.1$ & $0.2$ & -- & $0.2$ & $< 0.1$ & $0.2$ \\
         Palmes \textit{et al.} \cite{palmes2025Sr} 
         & 444779044.0 & $0.8$ & & & & & & \\
         Dubé \textit{et al.} (clock) \cite{dube2017absolute}  
         & 444779044.0954853 & $0.8\cdot 10^{-6}$ & & & & & & \\
         \hline
         
         $\mathrm{D}_{5/2}\rightarrow\mathrm{D}_{3/2}$ & & & & & & & & \\
         optical voltage calibration 
         & 8405351.8 & $1.5$ & $0.6$ & $0.8$ & $0.7$ & $0.3$ & $< 0.1$ & $0.8$ \\
         Palmes \textit{et al.} \cite{palmes2025Sr} 
         & 8405349.5 & $0.9$ & & & & & & \\
         \hline
         
         $\mathrm{S}_{1/2}\rightarrow\mathrm{D}_{5/2}$ & & & & & & & & \\
         optical voltage calibration 
         & 436373693.7 & $1.0$ & $0.2$ & $0.2$ & $0.6$ & $< 0.1$ & $< 0.1$ & $0.8$ \\
         Palmes \textit{et al.} \cite{palmes2025Sr} 
         & 436373694.6 & $0.9$ & & & & & & \\
         \hline\hline
    \end{tabular}
    \label{tab:FreqResults}
\end{table*}
\sisetup{input-symbols = {()},  
         group-digits  = false} 

\subsection{\label{secUncertainties}Uncertainty discussion}
Besides the uncertainties caused by the AC-Stark shift, which include uncertainties in $m_\mathrm{AC}$ as well as the detunings, the laser beam diameters and the laser powers, the main systematic uncertainties are the uncertainty in the laser frequency of the diode laser, the beam overlap, the ion energy, and the line shape. 

\subsubsection{\label{ssec:Lineshape}Line shape}
As discussed in Sec.\,\ref{ssec:LineShape}, the resonances display a strong asymmetry and substructure. To estimate the uncertainty introduced by this line shape, the fits were performed using the triple Gaussian, the asymmetric Voigt, and the thermal line shapes mentioned above. For the triple Gaussian, the center of mass, the tallest sub-peak, and the global maximum of the fit function were considered as the center. The uncertainty was estimated from the standard deviation of the difference in fit results. In the collinear-anticollinear approach, this cancels out in first order \cite{Krieger2017Be}, and the uncertainty is only $0.2\,$MHz. When using the optical calibration, this is not the case, as the line used as reference has a different line shape. Here, the uncertainty is $0.8\,$MHz.

\subsubsection{Laser Frequency}
\begin{figure}
    \centering
    \includegraphics[width=1\linewidth]{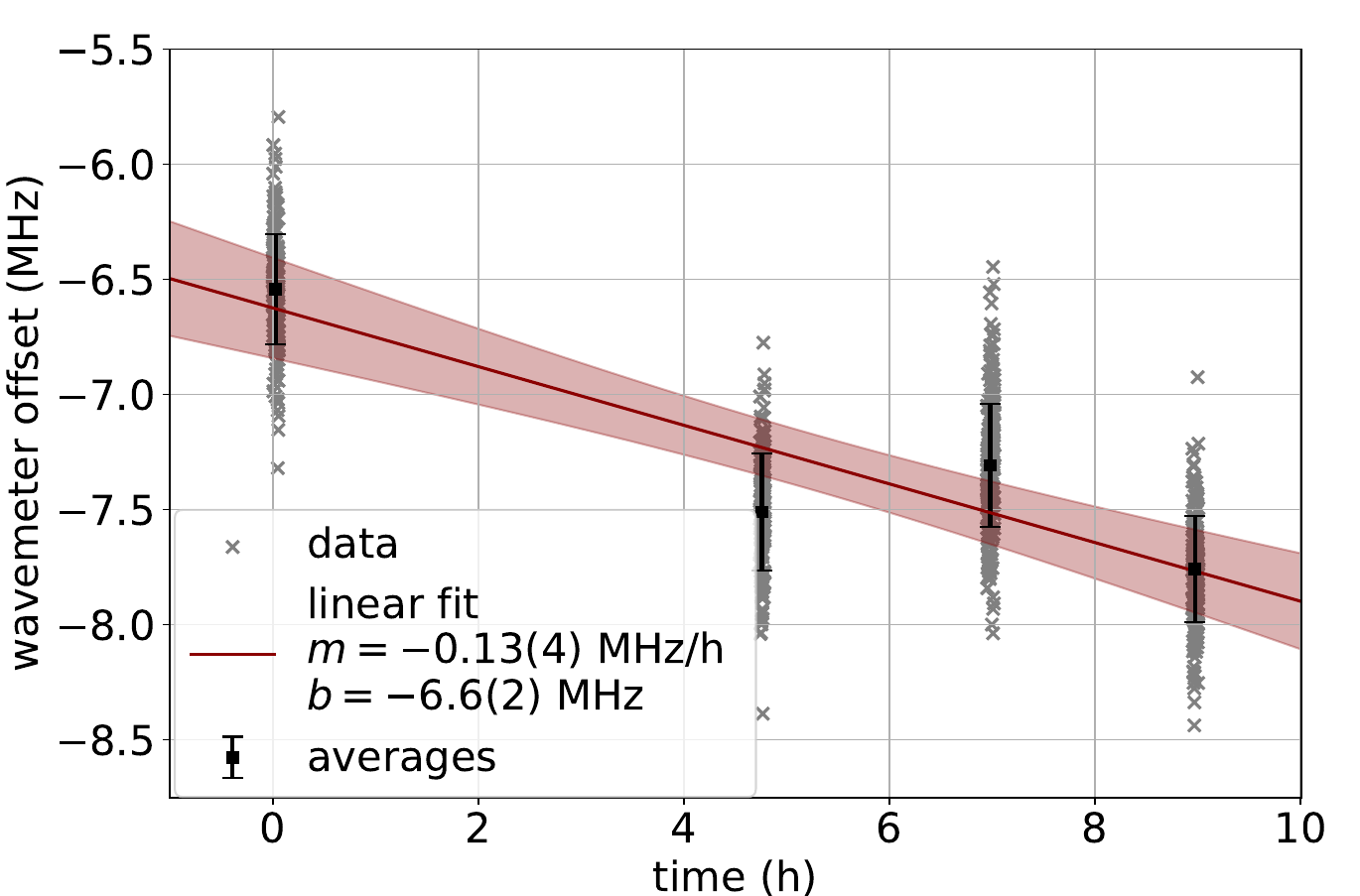}
    \caption{Measurements of the frequency offset $\Delta \nu = \nu_\mathrm{comb} - \nu_\mathrm{WS8-2}$ between the WS8-2 wavemeter ($\nu_\mathrm{WS8-2}$) used to monitor the diode laser and the frequency comb ($\nu_\mathrm{comb}$), taken over a span of 9 hours. The error bars indicate the averages and standard deviations of the individual measurements. A linear function $\Delta \nu=mt+b$ is then fitted to those averages (red line) to correct the frequency of the diode laser. The shaded area marks the confidence interval of the fit.}
    \label{fig:DiodeOffsetCal}
\end{figure}

Stabilizing the laser on the frequency comb, sub-$50\,$kHz long-term stability can be achieved. However, the frequency comb is designed to measure up to two laser frequencies in parallel. During measurements, the two Ti:Sa lasers are stabilized to the frequency comb, and the diode laser is stabilized to the WSU30 wavemeter, and its frequency is measured with the WS8-2 wavemeter. To correct for the offset inherent to this type of Fizeau interferometer-based wavemeter \cite{konig2020WavemeterPerformance}, measurements of the diode laser frequency, with the WS8-2 and the frequency comb in parallel, are performed between measurements. The difference between the frequencies measured with the wavemeter and the frequency comb is determined and interpolated to the individual frequency measurements, allowing a reduction of the systematic uncertainty in the diode laser frequency from $10\,$MHz to $300\,$kHz. An example of such an interpolation is shown in Fig.\,\ref{fig:DiodeOffsetCal}.
The uncertainty of $300\,$kHz is estimated based on the $1\sigma$ confidence interval of the interpolation and the uncertainty of the individual offset measurements. 

\subsubsection{Beam alignment}
The laser-laser overlap of all three lasers is optimized over a distance of $10\,$m, ensuring a misalignment of less than $0.2\,$mrad, and the ion-laser overlap is optimized on the multi-channel plates in the beam diagnostic stations, ensuring a misalignment of less than $0.4\,$mrad. For the investigated transition frequencies at the given beam energy and geometry, this results in an uncertainty of less than $14\,$kHz in the quasi-simultaneous measurements. For those performed using the optical voltage calibration, the uncertainties are estimated by sampling the different beam angles from uniform distributions with limits corresponding to the maximum expected angles. The resulting uncertainties are $13\,$kHz,  $24\,$kHz, and  $14\,$kHz for the $\mathrm{S}_{1/2}\rightarrow\mathrm{D}_{3/2}$, the $\mathrm{S}_{1/2}\rightarrow\mathrm{D}_{5/2}$, and the $\mathrm{D}_{3/2}\rightarrow\mathrm{D}_{5/2}$ transitions, respectively.

\subsubsection{Ion energy}
Given the available laser system, only the $\mathrm{S}_{1/2}\rightarrow\mathrm{D}_{5/2}$ Raman transition frequency can be measured using the quasi-simultaneous collinear-anticollinear approach. Thus, measuring the $\mathrm{S}_{1/2}\rightarrow\mathrm{D}_{3/2}$ and the $\mathrm{D}_{3/2}\rightarrow\mathrm{D}_{5/2}$ transitions requires precise knowledge of the effective acceleration voltage and, therefore, ion energy. 
The high voltage applied to the ion source is measured at a precision of $10^{-6}$ using a voltage divider calibrated by the PTB Braunschweig and a Keysight 3458 A $\mathrm{8\frac{1}{2}}$-digit multimeter.
The amplification factor of each Kepco voltage amplifier is determined by scanning the DAC input voltage while measuring the output voltage using an identical multimeter, allowing the determination of the total applied acceleration voltage $U_\mathrm{acc}-U_{\mathrm{IR} i}$ with a relative precision of $10^{-4}$.
However, the ion energy is also affected by space charge effects in the ion source as well as field penetrations and contact potentials. Thus, a laser spectroscopic voltage calibration was performed by measuring the $^{88}$Sr$^+$  D2-line as described in Sec.\,\ref{ssec:QuasiSim_freqs}. Comparison of the obtained voltages with the voltages measured electronically reveals an offset of $\sim 1\,$V. No dependency of this offset on the voltage $U_{\mathrm{IR}i}$ applied to the interaction region is found, indicating that it is predominantly caused by contact potentials, space charge effects, or voltage gradients in the ion source.

As measurements of the voltage offset differ over time due to, e.g., different conditions in the ion source, quasi-simultaneous optical voltage calibrations are performed during the measurements by including the D2-line in the measurements: Analogously to the quasi-simultaneous collinear-anticollinear measurements, the D2-line and the Raman transition of interest are measured in alternation. 
The electronically measured acceleration voltage of each Raman measurement is corrected by the difference between the optically and electronically measured voltages of a prior measurement of the D2-line.
As no dependence of the offset on $U_{\mathrm{IR}i}$ is observed, this is done without any correction concerning the difference in $U_{\mathrm{IR}i}$ at which the Raman resonance and the D2-line were recorded.
Instead, based on the scatter of the measured voltage offsets at different voltages applied to the interaction region, a systematic uncertainty of $0.05\,$V is introduced. This also covers statistical uncertainties in the optical voltage measurement and short-term fluctuations between the measurement of the D2-line and the Raman resonance. The resulting uncertainty in frequency space is of the same order of magnitude as the uncertainty induced by the $0.6$-MHz uncertainty of the D2 absolute transition frequency given in \cite{palmes2025Sr}, which has to be taken into account additionally.
We note that without the ion energy calibration, the obtained value would have been shifted by $\sim 20\,$MHz. This reflects the importance of the applied optical voltage calibration.

\section{\label{secOutlook}Outlook \& Conclusion}

It has been demonstrated that high-precision measurements of dipole-forbidden fine-structure transitions in CLS via Raman transitions are feasible. 
All transitions in the $^{88}$Sr$^+$ \Sone, \Pthree, $\mathrm{D}_\mathrm{3/2,5/2}$ $\Lambda$-scheme were measured. Those that could not be measured in collinear-anticollinear measurements were instead measured with an optical energy calibration via the \mbox{408-nm} D2-line.
Furthermore, a new scheme for performing Doppler-free Raman spectroscopy was used for the first direct $\mathrm{D}_{5/2}\rightarrow\mathrm{D}_{3/2}$ transition frequency measurement in $^{88}$Sr$^+$ and improved the accuracy of the so-far best determination, which was based on the difference of E1 transition frequencies in \cite{palmes2025Sr}.
Further improvements in precision could be achieved by investigating the AC-Stark shift with a frequency comb and by performing collinear-anticollinear measurements at a higher detuning of $1-2\,$GHz, which would further decrease the systematic uncertainty induced by the AC-Stark shift. 
The measurements performed on the $^{88}$Sr$^+$ $\mathrm{S}_{1/2}\rightarrow\mathrm{D}_{5/2}$ clock transition also yield excellent agreement with literature values.

In the future, this approach could be extended to isotopes with hyperfine structure, where Raman transitions have already been used to resolve hyperfine splitting too small to be investigated using dipole transitions \cite{dinneen1991stimulated}.
With the application of this scheme, it was also successfully demonstrated that populating metastable states via Raman transitions is feasible for subsequent high-precision collinear laser spectroscopy measurements and that this approach benefits from the small linewidth of the Raman transition, allowing the selective transfer of ions of a narrow velocity width \cite{Spahn2026RamanHV}.
However, further improvements in the stability of the interaction potential and measurements at a higher detuning are necessary to perform spectroscopy at the precision level of the intrinsic width of the Raman transition, and exploit this method to its full extent.
This level of precision would enable investigations of nuclear octupole moments and higher-order moments of the nuclear charge distribution, providing stringent tests of state-of-the-art atomic and nuclear theory \cite{yang2023laser, Koenig2026C14, PGR4thRadialMoment, deGroote2022Sc45, Bofos2024Octupole} and would allow contributing to searches for new bosons through King-plot nonlinearities extendable to short-lived nuclei \cite{Frugiuele.2017,Berengut.2018,Door.2025,Fuchs.2025,Wilzewski.2025}.

\section*{Data availability statement}
The data that support the findings of this study is openly available at the following URL/DOI: https://doi.org/10.48328/tudatalib-2037

\begin{acknowledgments}
We thank the group of Ruben de Groote from KU Leuven for providing the diode laser without which many of the performed measurements would not have been possible. We acknowledge support from the German Research Foundation (DFG, Deutsche Forschungsgemeinschaft) under project numbers NO789/4-1 and -- Project-ID 279384907 -- SFB 1245, as well as from the German Federal Ministry of Research, Technology and Space BMFTR under Contract Nos. 05P21RDFN1 and 05P24RD8. J.S. and H.B. acknowledge support from HGS-HIRe and R.V.D. from Fonds Wetenschappelijk Onderzoek Odysseus Project No. G0F7321N and KU Leuven grant C14/22/104. 
\end{acknowledgments}


\bibliography{SpahnRaman}

\end{document}